\PassOptionsToPackage{table}{xcolor}
\documentclass[sigconf, nonacm, pdfa, dvipsnames]{acmart}

\AtBeginDocument{%
  }

\newcommand\vldbdoi{10.14778/3705829.3705834}
\newcommand\vldbpages{130 - 143}
\newcommand\vldbvolume{18}
\newcommand\vldbissue{2}
\newcommand\vldbyear{2024}
\newcommand\vldbauthors{\authors}
\newcommand\vldbtitle{\shorttitle} 
\newcommand\vldbavailabilityurl{https://github.com/uwdb/color}
\newcommand\vldbpagestyle{empty}

\usepackage{amsthm}
\usepackage{amsmath}
\usepackage{amsfonts}
\usepackage{subcaption}
\usepackage{footmisc}
\usepackage{algorithm}
\usepackage{algpseudocode}
\usepackage{booktabs}

\usepackage[subtle]{savetrees}

\newtheorem{thm}{Theorem}

\newtheorem{corr}{Corollary}
\newtheorem{definition}{Definition}
\newtheorem{example}{Example}

\newcommand{\set}[1]{\{#1\}}                    
\newcommand{\setof}[2]{\{{#1}\mid{#2}\}}        

\newcommand{\dan}[1]{\textcolor{red}{{\textbf{Dan:} #1}}}

\newcommand{\rone}[1]{#1}

\newcommand{\rthree}[1]{#1}

\newcommand{\defeq}{\stackrel{\text{def}}{=}}
\newcommand{\Rp}{{\mathbb R}_{\tiny +}} 

\newcommand{\calC}{\mathcal C}
\newcommand{\calG}{\mathcal G}

\newcommand{\calP}{\mathcal P}

\newcommand{\eat}[1]{}
\usepackage[a-2b]{pdfx}

\begin{document}
\title{Color: A Framework for Applying Graph Coloring to Subgraph Cardinality Estimation}

\author{Kyle Deeds}
\thanks{The first two authors contributed equally to this work.}
\affiliation{%
  \institution{University of Washington}
  \streetaddress{}
  \postcode{}
}
\email{kdeeds@cs.washingon.edu}

\author{Diandre Sabale}
\affiliation{%
  \institution{University of Washington}
  \streetaddress{}
  \postcode{}
}
\email{dmbs@uw.edu}

\author{Moe Kayali}
\affiliation{%
  \institution{University of Washington}
  \streetaddress{}
  \postcode{}
}
\email{kayali@cs.washington.edu}

\author{Dan Suciu}
\affiliation{%
  \institution{University of Washington}
  \streetaddress{}
  \postcode{}
}
\email{suciu@cs.washington.edu}


\begin{abstract}
Graph workloads pose a particularly challenging problem for query optimizers. They typically feature large queries made up of entirely many-to-many joins with complex correlations. This puts significant stress on traditional cardinality estimation methods which generally see catastrophic errors when estimating the size of queries with only a handful of joins. To overcome this, we propose COLOR, a framework for subgraph cardinality estimation which applies insights from graph compression theory to produce a compact summary that captures the global topology of the data graph. Further, we identify several key optimizations that enable tractable estimation over this summary even for large query graphs. We then evaluate several designs within this framework and find that they improve accuracy by up to {$10^3\times$} over all competing methods while maintaining fast inference, a small memory footprint, efficient construction, and graceful degradation under updates.
\end{abstract}

\maketitle

\pagestyle{\vldbpagestyle}
\begingroup\small\noindent\raggedright\textbf{PVLDB Reference Format:}\\
\vldbauthors. \vldbtitle. PVLDB, \vldbvolume(\vldbissue): \vldbpages, \vldbyear.\\
\href{https://doi.org/\vldbdoi}{doi:\vldbdoi}
\endgroup
\begingroup
\renewcommand\thefootnote{}\footnote{\noindent
This work is licensed under the Creative Commons BY-NC-ND 4.0 International License. Visit \url{https://creativecommons.org/licenses/by-nc-nd/4.0/} to view a copy of this license. For any use beyond those covered by this license, obtain permission by emailing \href{mailto:info@vldb.org}{info@vldb.org}. Copyright is held by the owner/author(s). Publication rights licensed to the VLDB Endowment. \\
\raggedright Proceedings of the VLDB Endowment, Vol. \vldbvolume, No. \vldbissue\ %
ISSN 2150-8097. \\
\href{https://doi.org/\vldbdoi}{doi:\vldbdoi} \\
}\addtocounter{footnote}{-1}\endgroup

\ifdefempty{\vldbavailabilityurl}{}{
\vspace{.3cm}
\begingroup\small\noindent\raggedright\textbf{PVLDB Artifact Availability:}\\
The source code, data, and/or other artifacts have been made available at \url{\vldbavailabilityurl}.
\endgroup
}

\section{Introduction}
\label{sec:introduction}
The core operation of queries over graphs is \textit{subgraph matching} where instances of a query graph pattern, $Q$, are found within a larger data graph, $G$. This is the main primitive in graph query languages like GQL, SQL/PGQ, and SPARQL which are implemented by graph database management systems such as Neo4J, TigerGraph, Virtuoso, QLever, and Amazon Neptune ~\cite{Neo4J, DBLP:journals/corr/abs-1901-08248, DBLP:books/sp/virgilio09/ErlingM09, Bast:2017aa, DBLP:conf/semweb/BebeeCGGKKMMPRR18}. On critical workloads like financial fraud detection, subgraph matching is a part of virtually all analysis pipelines~\cite{DBLP:journals/vldb/SahuMSLO20}. For example, money laundering often manifests as cycles of transactions in financial networks~\cite{DBLP:journals/pvldb/QiuCQPZLZ18, DBLP:conf/adbis/HajduK20}.

Estimating the count, or \textit{cardinality}, of subgraphs is crucial to the optimization of graph queries. Specifically, subgraph matching algorithms are generally either search-based or join-based~\cite{sun2020memory, DBLP:journals/csur/BestaGPFPBAH24}. In both cases, the query optimizer attempts to choose a query plan (a query-vertex order or join order, respectively) which minimizes the size of intermediate results. However, graph workloads commonly contain large query graphs with ten or more edges; these queries may produce much larger intermediates which makes finding a good query plan particularly crucial for execution~\cite{DBLP:conf/www/0001SCBMN19, DBLP:journals/sigmod/MartensT19, DBLP:journals/vldb/BonifatiMT20}. Historically, cardinality estimation has been the key obstacle in identifying good query plans over relational data~\cite{DBLP:journals/pvldb/LeisGMBK015, DBLP:conf/sigmod/ParkKBKHH20, deeds2023safebound}. Applying these lessons to the graph setting, the database community has begun a concerted effort to improve subgraph cardinality estimation, with many recent methods and benchmarks~\cite{DBLP:conf/sigmod/ParkKBKHH20, DBLP:conf/icde/NeumannM11, DBLP:conf/sigmod/KimKFH21, DBLP:journals/tkdd/ChenL18, DBLP:conf/www/StefanoniMK18}. In this vein, our paper applies novel insights from graph theory to improve cardinality estimation for subgraph matching.


\begin{figure}
    \centering
    \includegraphics[width=\linewidth]{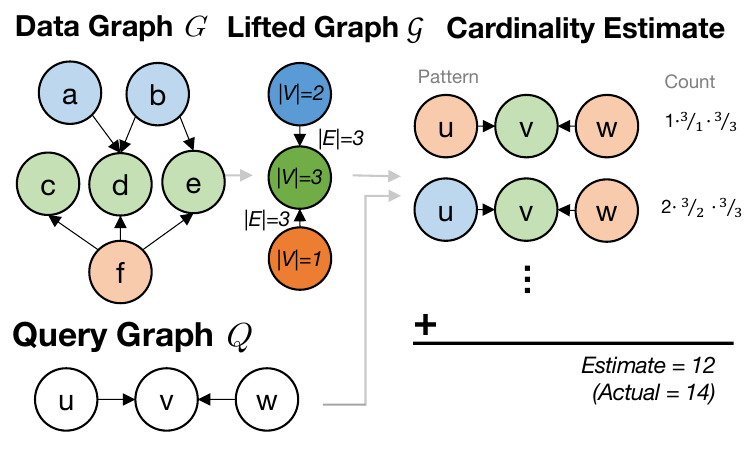}
    \caption{Lifted counting example. We estimate the number of occurrences of the query graph $Q$ pattern ($u \rightarrow v \leftarrow w$) in the data graph $G$. First, the data graph is partitioned offline: the resulting summary is stored as the \textit{lifted graph} $\mathcal{G}$. At runtime, the cardinality estimate is computed on the lifted graph $\mathcal{G}$ without reference to the underlying data graph.}
    \label{fig:intro-figure}
\end{figure}
Prior work on this problem generally follows one of three approaches. \textit{Pattern-based methods} \cite{DBLP:conf/icde/NeumannM11, wang2018presto} count the frequency of a small, pre-defined set of patterns offline, combining them at runtime to produce estimates by assuming independence between their counts. Simple queries can be effectively captured by the combination of one or two patterns---complex ones cannot, as they require combining many correlated patterns. For this reason, performance is adequate on simple queries but degrades rapidly for workloads with larger queries: resulting in median underestimates of up to $\mathbf{10^{10}\times}$ (Sec. ~\ref{sec:evaluation}). \textit{Online-sampling methods} \cite{DBLP:journals/tods/LiWYZ19, DBLP:journals/tkdd/ChenL18, DBLP:conf/sigmod/KimKFH21, DBLP:journals/pvldb/VengerovMZC15, DBLP:conf/sigmod/ZhaoC0HY18} perform runtime sampling, generally random walks, to estimate cardinalities. Unfortunately, these methods often suffer from sampling failure. While generating samples for smaller, less selective query patterns is feasible, finding even a single match for complex patterns can be challenging~\cite{liu2020understanding}. We later show $\mathbf{60-99\%}$ failure rate for online sampling methods on the most challenging workloads (Sec. ~\ref{sec:evaluation}). Further, online sampling in disk-based or distributed settings can incur a prohibitive latency as it relies on fast random reads over the whole graph.  \textit{Summarization methods} ~\cite{DBLP:conf/www/StefanoniMK18,DBLP:conf/sigmod/CaiBS19} group nodes in the data graph into a super structure and store summary statistics. However, the grouping is done by predefined rules or hashing, without regard to the edge distribution between these groups. Because these summaries are not tailored to the graph structure, \textit{i.e.} make a uniformity assumption, they produce median error of up to $\mathbf{10^{12}\times}$, and we find that they timeout on larger queries (Sec. ~\ref{sec:evaluation}).

\begin{figure}[t]
    \centering
    \includegraphics[width=\linewidth, keepaspectratio]{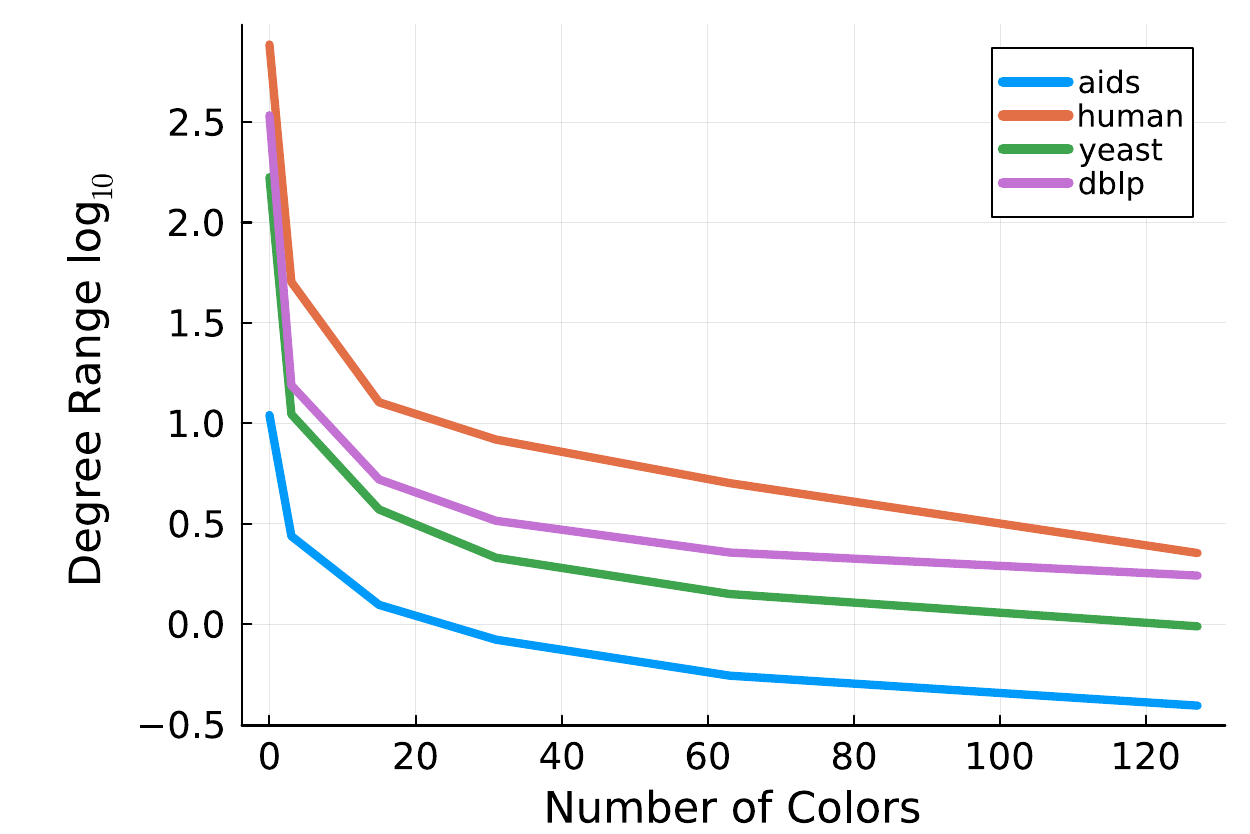}
    \caption{Accuracy of coloring as the number of colors increases} 
    \label{fig:degree-variance}
\end{figure}

In this paper, we propose a new approach to subgraph cardinality estimation based on graph compression; we call this method \textsc{Color}.  By taking advantage of recent advances in lossy graph representations such as quasi-stable coloring~\cite{DBLP:journals/pvldb/KayaliS22}, we approximately capture the topology of the graph in a small summary. We then directly estimate cardinalities on the summary without needing to access the data graph.

The key idea is to color the graph $G$ such that nodes of the same color have a similar number of edges to each other color. This mitigates the effect of the uniformity and independence assumptions. In Figure~\ref{fig:intro-figure} for example, the coloring assigns the $c, d, e$ nodes to the same green color. This is because green nodes have a similar number of incoming edges from blue and orange nodes, and no out edges. Meanwhile, $a, b$ are assigned to the blue color as they have a similar number of outgoing edges to green nodes and none to orange nodes. This helps mitigate the uniformity assumption because, within nodes of a fixed color, the edge distribution is nearly uniform. Further, because high and low degree nodes tend to be placed in different colors, correlations in the connections between them can be identified, mitigating the independence assumption.


Real world graphs are more complex, but it turns out that our approach can meaningfully capture their topology with a small number of colors, typically just 32 in our experiments. Figure~\ref{fig:degree-variance} shows the maximum difference in edge counts from nodes in one color to nodes in another, averaged over all pairs of colors for four of our benchmark datasets. Lower values indicate a better coloring and smaller differences between the two most different nodes in each color. A handful of colors is sufficient to capture most of the graph topology and reduce non-uniformity by 1-2 orders of magnitude.

With these colorings in mind, we return to the example in Figure~\ref{fig:intro-figure}. During the offline phase, our method takes the data graph, $G$, and produces a compact summary, $\calG$ called a {\em lifted graph}, with one super-node for each color (Sec. \ref{sec:colorings-and-lifted-graphs}). We keep statistics on the number and kinds of edges which pass between these colors. During the online phase, the cardinality estimate is computed on the lifted graph by performing a weighted version of subgraph counting which we call {\em lifted subgraph counting} (Sec. \ref{sec:lifted-subgraph-counting}). To extend this to cyclic queries, which occur frequently in graph databases, we also propose a technique based on a novel statistic called the {\em path closure probability} (Sec. \ref{sec:handling-cycles}).

To enable efficient inference on a more detailed, and therefore accurate, lifted graph, we propose three critical optimizations. Tree-decompositions and partial aggregation, introduced in Sec. \ref{subsec:partial-aggregation}, reduce the inference latency by over $\mathbf{100\times}$. Importance sampling and Thompson-Horowitz estimation over the lifted graph, the subject of Sec. \ref{subsec:sampling-techniques}, ensure a linear latency w.r.t. the size of the query while maintaining a \textbf{6x} lower error than a naive sampling approach. Lastly, we demonstrate how to maintain the lifted graph under updates in Sec. \ref{subsec:updates}, reducing the need to rebuild the summary by providing reasonable estimates even when over $1/2$ of the graph is updated.

In summary, we make the following contributions:
\begin{itemize}
    \item Develop the COLOR framework for producing lifted graph summaries from colorings (Sec. \ref{sec:colorings-and-lifted-graphs}) and evaluate six possible coloring schemes (Sec. \ref{sec:coloring-methods}).
    \item Define a general formula for performing inference over a lifted graph, show its optimality for acyclic queries (Sec. \ref{sec:lifted-subgraph-counting}), and extend it to cyclic ones (Sec. \ref{sec:handling-cycles}).
    \item Develop optimizations that allow for efficient and accurate inference and robust handling of updates (Sec. \ref{sec:optimization}). These optimizations are: tree-decomposition, importance sampling, and Thompson-Horowitz estimation.
    \item Empirically validate COLOR's superior performance on eight standard benchmark datasets and against nine comparison methods (Sec. \ref{sec:evaluation}).
\end{itemize}


\section{Problem Setting}

\label{sec:problem-setting}
\paragraph{Property Graphs} The data model that we use for this work is property graphs. These are directed graphs where each edge and vertex is associated with a set of attributes. These attributes can be simple labels (e.g. $:friendOf$) or key-value pairs (e.g. $Age=72$). This is the most general data model for graphs; it matches the model of GQL,  Cypher, and GraphQL, and it captures the RDF model \cite{francis2018cypher,hartig2018semantics,angles2017foundations}.

Formally, we define property graphs as follows:
\begin{definition}
    \label{def:labeled-graph}
    A property graph, $G(V, E, \lambda, \chi)$, is a directed graph with vertices $V$, edges $E$, attribute domain $\mathbb{A}$, and two annotation functions,
    \begin{itemize}
        \item $\lambda: \quad V\rightarrow 2^\mathbb{A}$ which maps a vertex to a set of attributes \footnote{$2^X$ denotes the set of subsets of X.}
        \item $\chi: \quad E \rightarrow 2^\mathbb{A}$ which maps an edge to a set of attributes
    \end{itemize}
\end{definition}

\paragraph{Subgraph Counting} The goal of subgraph counting is to find the number of occurrences of a query graph $Q$ in a larger data graph $G$. On a property-less graph, an occurrence is defined as any mapping from the vertices in $Q$ to the vertices in $G$ such that all edges in $Q$ are mapped to edges in $G$, i.e. a homomorphism from $Q$ to $G$. To account for properties, each vertex and edge of the query graph is associated with a predicate,  $P$, that returns true or false based on the attributes (e.g. "\textit{hasLabel:friendOf}" or "\textit{age>60}"). Formally, this is defined as follows,


\begin{definition}
    \label{def:subgraph-counting}
    For a property-less query graph $Q$ and data graph $G$, we define the set of subgraph matches as,
    \begin{align*}
        \hom(Q,G) = \setof{\pi: V_Q \rightarrow V_G}{\pi(E_Q) \subseteq E_G}
    \end{align*}
    Each match, $\pi$, is a function from $V_Q$ to $V_G$. When $Q$ and $G$ are property graphs, we add the natural conditions for each $\pi$,
    \begin{align}
       P_v(\pi(v)) = \mathbf{1} \quad\forall v \in V_Q \quad & \quad P_e(\pi(e)) = \mathbf{1} \quad\forall  e \in E_Q
    \end{align}
    The subgraph count is then $|\hom(Q,G)|$.
\end{definition}



This work studies the problem of \textit{cardinality estimation}. Recent work in both graph and relational databases has demonstrated the importance of cardinality estimation for producing efficient query plans \cite{park2020g, DBLP:journals/pvldb/LeisGMBK015}. This problem consists of two phases: 1) a preprocessing phase where the statistics, denoted $\mathbf{s}$, are computed and 2) an online phase where query graphs come in and approximate subgraph counts are returned. Formally, we can view it as follows,
\begin{definition}
    \label{def:cardinality-estimation}
    A cardinality estimation method $\mathcal{E}$ consists of two algorithms: 1) computing statistics during the preprocessing phase,  $\mathbf{s} \defeq \mathcal{E}_{pre}(G)$ $\,\,$ and 2) estimating the cardinality during the online phase, $c \defeq \mathcal{E}_{on}(Q, \mathbf{s}) \in \Rp$. The goal is to produce an estimate where $c \approx |\hom(Q,G)|$.
\end{definition}
The primary metrics for these algorithms are: 1) the accuracy of $c \approx |\hom(Q,G)|$ 2) the latency of $\mathcal{E}_{on}$ and 3) the size of $\mathbf{s}$.

\paragraph{Traditional Estimators} The classic System R approach to cardinality estimation in relational databases combines the number tuples in the joining relations, the number of unique values in the joining columns, and assumptions (uniformity, independence, preservation of values) to produce a basic cardinality estimate \cite{DBLP:journals/pvldb/LeisGMBK015, DBLP:journals/dr/Haas99a}. In the graph setting, this estimation method looks like the following,
\begin{definition}\label{def:traditional:estimator}
Given a query graph $Q(V_Q, E_Q)$ and data graph $G(V, E)$, the traditional estimation method is,
\begin{enumerate}
    \item $\mathcal{E}_{pre}^{trad}(G) = (|V|, |E|)$ (number of vertices and of edges)
    \item $\mathcal{E}_{on}^{trad}(Q, \mathbf{s}) = \prod_{v\in V_Q} |V| \cdot \prod_{e\in E_Q} \frac{|E|}{|V|^2} $
\end{enumerate}
\label{def:traditional-estimator}
\end{definition}

Intuitively, the estimation formula calculates the number of possible embeddings of the query graph in the data graph, $\prod_{v\in V_Q}|V|$, then scales this by the probability of any embedding having the correct set of edges, $\prod_{e\in E_Q}\frac{|E|}{|V|^2}$. In effect, this estimation procedure assumes that the data graph is distributed like an Erdos-Renyi random graph with $|E|$ edges and $|V|$ vertices and produces an accurate estimate given this assumption. However, the structure of most real world graphs is much more complex. This results in very different subgraph counts from those on Erdos-Renyi random graphs and motivates the use of more complex estimators.
\begin{example} \label{ex:traditional} Recall that the standard estimate of a join $Q(x,y,z) = R(x,y)\wedge S(y,z)$ is $\frac{|R|\cdot|S|}{\max(|R.y|,|S.y|)}$.  When both $R,S$ are the edge relation $E$, then $R.y=S.y=V$ (assuming no isolated vertices) and the traditional estimator becomes $\frac{|E|\cdot |E|}{|V|}$, which is the same as the formula above, $|V|^3 \frac{|E|^2}{|V|^4}$.
\end{example}
\section{Colorings \& Lifted Graphs}
\label{sec:colorings-and-lifted-graphs}
Colorings and lifted graphs are the core of our framework, so we start by formally defining them here. For clarity of presentation, we will begin by ignoring predicates and reintroduce them later. Given a graph $G(V,E)$, a {\em coloring} $\sigma$ is a function from $V$ to $C$ where $C$ is a small set of colors, $|C| \ll |V|$. Under a {\em quasi-stable} coloring~\cite{DBLP:journals/pvldb/KayaliS22}, two vertices in the same color will have similar distributions of outgoing edges to different colors, i.e. any two \textcolor{red}{red} vertices should have nearly the same number of edges to \textcolor{blue}{blue} vertices. Formally,
\begin{definition}
\label{def:good-coloring}
A coloring, $\sigma$, is quasi-stable if the following properties hold for all pairs of vertices $v_1, v_2$. If $\sigma(v_1)=\sigma(v_2)$, then:
    \begin{align}
\forall \,c\in C: |\{(v_1, v)\in E | \sigma(v) = c\}| \approx & |\{(v_2, v)\in E | \sigma(v) = c\}| \label{eq:quasi:stable}
    \end{align}
\end{definition}

In English, $\sigma$ is quasi-stable if for any two colors $c_0, c$, any two vertices $v_1,v_2$ colored $c_0$ have approximately the same number of neighbors colored $c$. If we replace $\approx$ with $=$ in~\eqref{eq:quasi:stable}, then $\sigma$ is called a {\em stable coloring}.  Stable colorings are commonly used in graph isomorphism algorithms, and have elegant theoretical properties~\cite{DBLP:books/cu/G2017, DBLP:journals/cacm/GroheS20}. However, they are unsuitable for our purpose, because stable colorings of real-world graphs require a very large number of colors~\cite{DBLP:journals/pvldb/KayaliS22}.  In fact, in a random graph, every vertex has a distinct color with high probability ~\cite{DBLP:journals/corr/abs-2011-01366}. Instead, we relax equality $=$ to approximation $\approx$ in Definition~\ref{def:good-coloring} in exchange for using a much smaller number of colors. To do this, we apply a variety of coloring algorithms (Sec. \ref{sec:coloring-methods}) which produce a dramatic reduction in the number of colors with only a small relaxation of $=$ to $\approx$. We demonstrate this in Fig. \ref{fig:degree-variance}. This graph shows the average range of degrees from nodes in one color to nodes in another as the number of colors varies. Across all graphs, a coloring with just 32 colors (using the Quasi-Stable coloring method from \cite{DBLP:journals/pvldb/KayaliS22}) lowers the average degree range by over 2 orders of magnitude as compared to the initial graph.



For any color $c \in C$, we denote the set of vertices colored $c$ by $V_c \defeq \setof{v \in V}{\sigma(v)=c}$. For any two colors $c_1, c_2$ we denote the set of edges between them by $E_{c_1c_2} \defeq E \cap (V_{c_1} \times V_{c_2})$.  The \textit{lifted graph} consists of a quasi-stable coloring, together with statistics for each pair of colors:

\begin{definition}
    \label{def:lifted-graph}
    Fix a directed graph $G=(V,E)$, and a coloring $\sigma$ with colors $C$.  A \emph{lifted graph} is a triple, $\calG = (F, \psi, \tau)$, consisting of the following parts.
    \begin{itemize}
    \item $F = (V_F, E_F)$ is a graph where $V_F = C$ and $E_F = \setof{(c_1,c_2)}{E_{c_1c_2}\neq\emptyset}$.
    \item  $\psi: C \rightarrow \mathbb{N}$ where $\psi(c) = |V_c|$
    \item $\tau:E_{c_1c_2} \rightarrow \Rp$ maps edges of the graph to statistics about the edges in $E_{c_1c_2}$
    \end{itemize}
\end{definition}

In this paper, we consider the following choices for the edge statistics function $\tau$:
\begin{align*}
  \tau_{\text{min}}(c_1,c_2) = & \min \setof{|\setof{(v_1,v_2)}{v_2 \in V_{c_2}}|}{v_1\in V_{c_1}}\\
  \tau_{\text{avg}}(c_1,c_2) = & \frac{|E_{c_1c_2}|}{|V_{c_1}|}\\
  \tau_{\text{max}}(c_1,c_2) = & \max \setof{|\setof{(v_1,v_2)}{v_2 \in V_{c_2}}|}{v_1\in V_{c_1}}
\end{align*}
They represent the minimum degree, the average degree, and the maximum degree of a vertex in $c_1$ to vertices in $c_2$ respectively.  In the following sections, it will be sufficient to assume that $\tau$ means $\tau_{\text{avg}}$, unless otherwise noted.

%

Thus, $\psi$ is a function that returns the number of vertices with the color $c$, and $\tau$ is a function that returns statistics about the set of edges between a pair of colors.  The lifted graph forms a fuzzy compression of the data graph, and is computed offline, during preprocessing. In our approach, this is the statistic, $\mathbf{s}=\mathcal{E}_{pre}$, as defined in Def. \ref{def:cardinality-estimation}.

\begin{example}
Consider again the example data graph and coloring in Figure \ref{fig:intro-figure}. First, the data graph $G$ is colored with a quasi-stable coloring. This produces three different colors, which reflect that topologically there are three different "kinds" of vertices in the data graph. In particular, orange vertices have an 3 out-degree, no in-degree; the green vertices no out-degree, 2 out-degree; and blue 1.5 average out-degree, no in-degree. 
Due to the arrangement of these edges, we can produce a ``good'' coloring where vertices $a$ and $b$ are assigned to the blue partition. Further, we can produce the lifted graph $\mathcal{G}$ by choosing the degree sum as our edge statistic. In this figure, the vertex labels are the partition cardinalities, i.e. the values of $\psi$. The label edges represent the sum of edges between two colors: for example, green vertices have a total of 3 edges into the orange vertices, i.e. the values of $\tau_\Sigma$. Note that this lifted graph closely captures the distribution of edges in the data graph while being half the size.
\end{example}

\paragraph{Lifted Property Graphs} 
To account for attributes and predicates in the lifted graph, we adjust the definition of $\psi$ and $\tau$ to accept predicates in addition to colors. Suppose that a query has a vertex predicate $P$, then we define $V_{c,P}$ as the set of data vertices in the color $c$ which pass the predicate $P$. Similarly, $E_{c_1,c_2,P_e, P_v}$ is the set of edges starting in color $c_1$, matching predicate $P_e$, and landing on a node in color $c_2$ which matches $P_v$. With this, we can then redefine $\psi$ and $\tau_{avg}$,
\begin{align*}
  \psi(c, P) = & |V_{c,p}| & \tau_{avg}(c_1,c_2,P_e,P_v) = & \frac{|E_{c_1,c_2,P_e,P_v}|}{|V_{c_1}|}
\end{align*}
We allow these functions to be exact or approximate in order to accommodate more complex predicates. If the predicates are all of the form $hasLabel:X$ and there are few edge label/vertex label combinations, then this can be calculated and stored explicitly. However, predicates like range or LIKE benefit from approximating $|E_{c_1,c_2,P_e,P_v}|$ using standard techniques like histograms or n-grams. This can then be extended to $t_{min}$ and $t_{max}$ using techniques similar to those in \cite{deeds2023safebound}.
\begin{table}[]
\caption{Notation Dictionary}
\begin{tabular}{lp{2.5in}}
\toprule
\footnotesize
\textbf{Symbol}       & \textbf{Meaning}                                                                                               \\ \midrule
$G(V,E,\lambda,\chi)$ & Property graph with vertices $V$, edges $E$, vertex attributes $\lambda$, 
and edge attributes $\chi$.          \\ 
$\calG(F,\psi,\tau)$  & Lifted graph with color graph $F$, color cardinalities $\psi$,
and color edge statistics $\tau$.               \\
$W(\pi, Q, \calG)$              & Estimate of the subgraph count for query $Q$ with coloring $\pi$ based on lifted graph $\calG$. \\ 
$\Phi(Q,\calG)$       & Estimate of the subgraph count for query $Q$ based on lifted graph $\calG$.\\
$\gamma(C_1,C_2,D)$              & Probability of a path closing a cycle from $C_1$ to $C_2$ with directionality $D$                                                                                     \\ \bottomrule
\end{tabular}
\end{table}
\section{Lifted Subgraph Counting}
\label{sec:lifted-subgraph-counting}
We have described the lifted graph, a small weighted graph which approximately captures the topology of the data graph.  Next, we show how to use the lifted graph for cardinality estimation, which we call  the \textit{lifted subgraph counting problem},
\begin{definition}
    \label{def:lifted-subgraph-counting}
    Fix a lifted graph $\calG = (F, \psi, \tau)$ and a query graph $Q$.  An estimation procedure is a function
    \begin{align*}
      W:  \hom(Q,F)\times Q\times \calG \rightarrow \Rp
    \end{align*}
    The lifted subgraph count is,
    \begin{align}
      \Phi(Q, \calG) = \sum_{\pi\in \hom(Q,F)} W(\pi, Q, \calG)  \label{eq:method:lifted-subgraph-counting}
    \end{align}
\end{definition}

A homomorphism $\pi:Q \rightarrow F$ associates to each query vertex $x$ a color $\pi(x)\in C$; we will also call $\pi$ a \emph{coloring} of the query $Q$.  The estimator $W(\pi, Q, \calG)$ approximates the number of outputs of the query with coloring $\pi$. Note, we generally drop $Q$ and $\calG$ when obvious from context. The total estimate, $\Phi(Q, \calG)$ is simply the sum over all colorings $\pi$.  To see the intuition, recall that errors in traditional cardinality estimation come from correlation and skew. For example, the former could mean that high degree vertices are more likely to be connected to high degree vertices (or vice versa), while the latter means that the degrees of vertices have a wide range. By grouping vertices into colors based on their local topology (their degrees, their neighbors' degrees, etc), and fixing a particular coloring $\pi$ of the query vertices, we reduce the variance of the estimate $W(\pi)$, leading to a reduced error overall.  In the rest of this section we define the estimate function $W$ assuming that the query graph is acyclic.  We will extend it to arbitrary query graphs in Sec.~\ref{sec:handling-cycles}.


\subsection{Acyclic Query Graphs }


For acyclic queries, we use the following function $W$:

\begin{definition}[Lifted Estimator for Acyclic Queries]
    \label{def:standard-estimator}
    Let $Q=(V_Q,E_Q)$ be an acyclic query graph, and let $x_1,\ldots,x_{|V_Q|}$ be a topological ordering of its vertices: in other words, every vertex $x_j$ is connected to some $x_i$ for $i<j$.  For any homomorphism $\pi : Q\rightarrow F$, we define:
    \begin{align}
        W(\pi) &\defeq \psi(\pi(x_1),P_{x_1}) \prod_{(x_i, x_j)\in E_Q}\tau(\pi(x_i), \pi(x_j), P_{x_j},P_{(x_i,x_j)})\label{eq:standard-estimator}
    \end{align}
\end{definition}
We adopt the convention that if $x_i$ and $x_j$ are connected by a reverse edge, i.e. $E_Q$ contains $(x_j,x_i)$ rather than $(x_i,x_j)$, then this is reflected in the query graph with a predicate, $dir=\leftarrow$, on the edge. Intuitively, we process the edges in the topological order, so we always multiply with the in/outdegree of the color assigned to the topologically earlier vertex. In this paper we choose the topological order such as to minimize the time needed to compute $\hom(Q,G_F)$, see Sec.~\ref{sec:optimization}.

\begin{example}
  Here we show that the lifted graph estimator generalizes the traditional estimator in Def.~\ref{def:traditional:estimator}. We illustrate this with a graph $G=(V,E)$ and the 2-edge query $Q(x,y,z) = E(x,y)\wedge E(y,z)$ from Example~\ref{ex:traditional}.  Assume that we use a lifted graph, $\mathcal{G}(F,\psi,\tau)$, consisting of a single color $c$ and a single edge: $F = (\set{c}, \set{(c,c)})$.  Then the statistics are $\psi(c)=|V|$ and $\tau(c,c) = \frac{|E|}{|V|}$ (recall that we assumed $\tau$ refers to $\tau_{avg}$), there is a single homomorphism $\pi:G\rightarrow F$, and our estimate is $W(\pi) = \psi(c) \left(\tau(c,c)\right)^2 = |V| \frac{|E|^2}{|V|^2} = \frac{|E|^2}{|V|}$.  This is the same as the traditional estimator in Example~\ref{ex:traditional}.
\end{example}

\begin{example} We illustrate now how a better designed lifted graph can lead to an improved estimator.  Assume that the data graph is the disjoint union of two graphs, $G=G_1 \cup G_2$, where $G_1(V_1, E_1)$ is a $2$-regular graph on 10000 vertices, and $G_2(V_2, E_2)$ is a clique of size 100.  Suppose $Q$ is a path of length $k$.  Since the average degree in $G$ is $\approx 4$, a traditional estimate for $Q$ is $10100 \cdot 4^k$, which vastly underestimates the true count, because it doesn't capture the skew and correlation introduced by the clique sub-graph.  Suppose we pre-compute a lifted graph consisting of two colors: \textcolor{OliveGreen}{green} contains all vertices $V_1$ and \textcolor{red}{red} color contains all vertices $V_2$.  There are only two edges $(\text{\textcolor{OliveGreen}{green}},\text{\textcolor{OliveGreen}{green}})$ and $(\text{\textcolor{red}{red}},\text{\textcolor{red}{red}})$, and therefore only two colorings $\pi$ of the query $Q$.  We compute $W$ separately for each of the two colorings, then return their sum, $100\cdot 99^k + 10000\cdot2^k$, which, for our simple data graph, is an exact count.
\end{example}

\subsection{Special Case: Stable Colorings}
\label{subsec:stable-colorings}
As a theoretical justification of our method, we prove that if the lifted graph is a {\em stable} coloring (meaning: $\approx$ is $=$ in Def.~\ref{def:good-coloring}), then our estimate for acyclic queries is exact, although we defer the formal proof to the technical report ~\cite{tech-report} for space.

\begin{thm}
  \label{thm:stable-colorings} Let $\calG$ be a lifted graph defined by a stable coloring $\sigma$.  Then $\tau_{\text{min}}=\tau_{\text{avg}}=\tau_{\text{max}}$, and, for any acyclic query $Q$, the lifted graph estimator is exact:
\begin{align}
  |\hom(Q,G)| = \Phi(Q, \calG) \label{eq:stable_coloring:Hom}
\end{align}
\end{thm}
This theorem states that stable colorings are a perfect statistic for cardinality estimation of acyclic query graphs.  However, we cannot use them in practice, because the number of stable colors needed to represent real graphs is close to the number of vertices in the data graph~\cite{DBLP:journals/corr/abs-2112-09992,DBLP:journals/pvldb/KayaliS22}. \rone{Fortunately, we can prove an additional corollary for quasi-stable colorings:}
\rone{
\begin{corr}
    Let $\epsilon$ be the approximation error of the lifted graph $\calG$ defined as,
    \begin{align}
        \epsilon = \max_{c_1,c_2\in C} \frac{\tau_{\max}(c_1,c_2)}{\tau_{\min}(c_1,c_2)}
    \end{align}
    Then, we can bound the error of our subgraph estimator as,
    \begin{align}
        max(\frac{\Phi(Q,\calG)}{|hom(Q,G)|},\frac{|hom(Q,G)|}{\Phi(Q,\calG)})\leq \epsilon^{(|V_Q|-1)}
    \end{align}
\end{corr}
Intuitively, as colorings approach stability, the estimate converges to the true subgraph count.
}
\section{Handling Cycles}
\label{sec:handling-cycles}
Cyclic queries are a fundamentally different challenge for cardinality estimators because they allow complex dependencies between vertices within the query graph. Practically, they require estimating the probability that an edge between two vertices exists, conditioned on the fact the these vertices are already connected in the query graph. This section outlines new techniques to estimate this {\em cycle closure probability}.

\subsection{Cycle Closure Probability}
\label{subsec:cycle-closure}
To ground this discussion, we begin by defining the probability space and random variables. The former is defined by a uniform random selection of $|V_Q|$ vertices from $V_G$ with replacement. This is equivalent to a random mapping from $V_G$ to $V_Q$. The set of vertices selected by this process is denoted with the random variables $V_1,\ldots,V_{|V_Q|}$. Further, each edge of the query graph, $(v_i, v_j) = e_i\in E_Q$, is associated with a binary random variable $E_i$ which is true when $(V_i, V_j)\in E_G$. In other words, $E_i$ is true iff the data vertices mapped to that edge of the query have an edge between them.

As a basic example, we can calculate the unconditioned probability of $E_i$ for any edge $e_i$ as follows,
\begin{align*}
    P(E_i) = \frac{|E_G|}{|V_G|^2}
\end{align*}
Further, we can express the cardinality of an arbitrary query as,
\begin{align*}
    |Hom(Q,G)| = |V_G|^{|V_Q|}\cdot P(\cap_{e_i\in E_Q}E_i)
\end{align*}
With conditional probability, we can expand the probability as,
\begin{align*}
    P(\cap_{e_i\in E_Q}E_i) = \prod_{e_i\in E_Q} P(E_i|E_1,\ldots,E_{i-1})
\end{align*}
When the endpoints of an edge are contained within the previous edges, $e_i\in\cap_{j=1}^{i-1}e_j$, the probability within the product is a cycle closure probability. It is this probability which we try to estimate in this section, and the crucial challenge is estimating the effect of the previous edges,  $E_1,\ldots,E_{i-1}$. 

The naive solution is to consider all patterns of a fixed size (i.e. the pattern induced by $E_1,\ldots,E_{i-1}$) and calculate the probability of an edge occurring between two nodes of that pattern in the data graph. However, the number of patterns increases super-exponentially in their size due to the choices of basic graph pattern, edge direction, and predicates, so this approach is infeasible for all but the smallest queries. Our approach attempts to tackle this by considering a smaller set of patterns and composing them smartly.

\subsection{Path Closure Probability}
We introduce {\em path closure probabilities} which represent the probability that a path in the data graph is "closed", i.e. there is an edge from the starting vertex to the ending vertex. To limit the number of probabilities that we store, paths are grouped by their directionality, e.g. $D = \{\leftarrow, \rightarrow\}^k$, and by the color of their starting and ending vertices. We denote this probability as,
\begin{definition}
    Let $\mathcal{P}_{c_1,c_2,D}(G)$ be the set of paths in the data graph $G$ with directions matching $D$ and starting/ending color $c_1$/$c_2$. Let $\mathcal{C}_{c_1,c_2,D}(G)$ be the subset of paths in $\mathcal{P}_{c_1,c_2,D}(G)$ which are closed. The path closure probability is then,
    \begin{align*}
    \gamma(c_1, c_2, D) = \frac{|\mathcal{C}_{c_1,c_2,D}(G)|}{|\mathcal{P}_{c_1,c_2,D}(G)|}
    \end{align*}
\end{definition}
Given this statistic, we can define an expression for the cycle closure probability, $P(E_i|E_1,\ldots,E_{i-1})$. Let $\mathcal{S}_{i-1}$ be the set of simple paths in $E_1,\ldots,E_{i-1}$ which start at the source of $E_i$ and end at its destination. Note that the closure of any path within $S$ implies that $E_i$ is true. Based on this, we treat the closure of each of these paths as an independent event (a conservative assumption), and calculate the cycle closure probability as,
\begin{align*}
    P(E_i|E_1,\ldots,E_{i-1}) = 1 - \prod_{(c_1,c_2,D)\in \mathcal{S}_{i-1}}(1-\gamma(c_1,c_2,D))
\end{align*}

We can now explain how we extend the definition of the acyclic estimator~\eqref{eq:standard-estimator} to handle arbitrary query graphs.  Fix a query graph $Q=(V_Q, E_Q)$, and consider a topological edge ordering, $e_1, \ldots, e_{|E_Q|}$, which means that every edge $e_j$ has a vertex in common with some previous edge $e_i$, $i<j$. This ordering defines a spanning tree $T$, consisting of the subset of edges that introduce a new vertex, i.e. $T = \setof{e_j}{e_j \not\subseteq \bigcup_{i<j} e_i}$.  If $e_i=(x,y)$ is not a tree edge, then both vertices $x,y$ are already connected in the subgraph consisting of $e_1, \ldots, e_{i-1}$. The modified definition of the estimator~\eqref{eq:standard-estimator} is:
\begin{align}
\label{eq:W:cyclic}
  W(\pi) &= \psi(\pi(x_1, \lambda_Q(x_1)))\prod_{e_i \in E_Q}\omega(e_i)\\
  \omega(e_i)
         &=
           \begin{cases}
             \tau(\pi(x),\pi(y), \lambda_Q(x),\chi_Q(x,y)))&\text{if }e=(x,y)\in T\\
            1 - \prod_{(c_1,c_2,D)\in \mathcal{S}_{i-1}}(1-\gamma(c_1,c_2,D))&\text{if }e=(x,y)\not\in T
           \end{cases}
\end{align}
To keep the construction of our statistics tractable, we do not calculate $\gamma$ exactly. Instead, we sample paths from the lifted graph and use these to calculate probabilities. As a default, we use $100,000$ sampled paths when calculating these statistics in our experiments. If a particular color combination doesn't occur in our samples, we fall back to the probability just conditioned on the sequence of directions.

\section{Alternate Coloring Methods}
\label{sec:coloring-methods}
Because a coloring can be any mapping from vertices to colors, there is a wide design space of algorithms for creating colorings. The goal of this section is to find colorings which facilitate improved cardinality estimations. In this work, we focus on divisive colorings where all vertices begin in the same color and then the following steps proceed iteratively: 1) identify a color, $c$, to split into two colors 2) for each vertex in $c$, determine whether it should stay in $c$ or join the new color. The benefit of this approach is that arbitrary coloring methods can be composed, allowing for more robustness and accuracy.

\paragraph{Quasi-Stable Coloring~\cite{DBLP:journals/pvldb/KayaliS22}} As explained earlier, this is a generalization of the traditional color-refinement algorithm. Rather than producing a stable coloring, this algorithm softens the requirements and instead requires vertices in each color to have a "similar" number of edges to each other color. At each iteration, it selects the color with the widest range of degrees w.r.t. another color and splits it into two colors with more uniform counts relative to the other color.

\paragraph{Degree Coloring} This coloring simply separates vertices into colors based on their overall degree. The intuition is that vertices with high degree are generally occupying similar positions in the data graph and vice versa with low degree vertices. It begins by selecting the color with the largest range of degrees to split, then separates vertices into two colors depending on whether they are above or below the average degree.

\paragraph{Neighbor Label Coloring} An alternative to quasi-stable coloring, this method attempts to reduce the variance introduced by label predicates (e.g. "hasLabel:X") by grouping vertices based on their neighbors' label attributes. First, it selects the color whose vertices have the widest range of degrees w.r.t. the vertex labels of their neighbors. It then splits it into two colors which have more uniform connections to vertices with each vertex label.

\paragraph{Vertex Label Coloring} A more direct version of the previous approach, this coloring also aims to reduce the effect of label predicates. This time it aims to make the distribution of vertex labels within colors more uniform. To do this, it first identifies the color, $c_1$, with the most even distribution of a particular label, weighted by size. The nodes in $c_1$ which have that label attribute are then put in a new color and the ones which do not remain in $c_1$.

\paragraph{Mixture Coloring} The previous colorings generally target a particular source of variance related to either topology or attribute distributions, and they divide the color where this kind of variance appears most strongly. So, it makes sense to layer these colorings in order to jointly manage these different concerns, and we call this a mixture coloring. In the experiments (Fig. \ref{fig:colorings-error}), we show that this is the most accurate coloring across a range of workloads.

\paragraph{Hash Coloring} For completeness, we consider the naive hash coloring which uniformly randomly sorts vertices into colors. This corresponds to the partitioning used by \cite{DBLP:conf/sigmod/CaiBS19} to tighten their cardinality bounds. This method is convenient because construction is linear in the size of $|G|$, and it does not require coordination in a distributed setting. However, it offers limited improvement to the estimator because it doesn't take the specific topology or attributes of the graph into account.

\begin{algorithm}
    \small
    \caption{Optimized Inference Algorithm}
    \begin{algorithmic}[1]
    \Require $F_{G,\sigma, l}(G_F(C,E_V), \psi, \tau), \,\,\,Q(V_Q, E_Q), \,\,\,v_1,\ldots,v_{|V_Q|}\quad\quad\quad$ // Lifted Graph, Query Graph, Vertex Order
    \State $PC = \{(\{\}, 1)\}$ // Partial Colorings
    \State $V_F = \{\}$
    \For{$i\in[1,\ldots,|Q|]$}
        \State $PC' = \{\}$
        \State $E_i = \{e \in E_Q\,\,|\,\, v_i\in e\}$
        \For{$\pi, w \,\in \,PC$}
            \For{$c \,\in\,C$}
                \State $\pi' = \pi \cup (v_i\rightarrow c)$
                \State $w' = w\cdot\prod_{e\in E_i}\omega(\pi', e)$ $\leftarrow$  Estimator Sec. \ref{subsec:cycle-closure}
                \State $PC' = PC' \cup \{\pi', w'\}$
            \EndFor
        \EndFor
        \State $PC = PC'$
        \State $V_{S} = \{v\in V| (v,v_j),(v_j, v) \not\in E_Q \,\,\forall\,\, j > i\} \setminus V_F$
        \State $V_F = V_F \bigcup V_S$
        \State $PC = \sum_{V_{S}} PC$ $\leftarrow$ Partial Aggregation Sec. \ref{subsec:partial-aggregation}
        \State $PC = Sample(PC)$ $\leftarrow$ Sampling Sec. \ref{subsec:sampling-techniques}
    \EndFor\\
    \Return $PC$
    \end{algorithmic}
    \label{alg:optimized-inference}
\end{algorithm}

\section{Optimization}
\label{sec:optimization}
\subsection{Partial Aggregation}
\label{subsec:partial-aggregation}
Naively, the runtime of inference on the lifted graph is exponential in the size of the query graph, making it intractable for even moderately sized query graphs. The lifted graph is dense so the size of $Hom(Q,F)$ is approximately $|C|^{|Q|}$. Fortunately, we can rephrase Eq. \ref{eq:method:lifted-subgraph-counting} to drastically reduce this runtime via aggregate push-down.  To do this, we express the set of lifted graph matches, $\text{hom}(Q,F)$, as $(c_1,\ldots,c_{|Q|})\in C^{|Q|}$. We then use the definition of $W$ from \ref{eq:W:cyclic} and, as before, we assume a topological ordering on the edges, $e_1,\ldots,e_{|E|}$, and vertices, $v_1,\ldots,v_{|Q|}$.
\begin{align}
        \Phi(Q, F, W_{\psi, \tau}) &= \sum_{c_1,...,c_{|Q|}\in C} \psi(c_1)\prod_{i=1}^{|E_Q|}\omega(\pi_{c_1, \ldots, c_{|Q|}},e_i) \label{eq:optimization:faq_version}
\end{align}
At this point, we can identify this as a {\em{Functional Aggregate Query}} \cite{DBLP:conf/pods/KhamisNR16} and apply the techniques there to solve it efficiently.\footnote{Note, this is closely related to the variable elimination algorithm for probabilistic graphical models as well as tensor contraction algorithms.} As an example, suppose that the query graph $Q$ is a line graph $v_1\rightarrow v_2\rightarrow v_3$. The naive expression would be the following $O(|C|^3)$ expression,
\begin{align*}
        \Phi(Q, F, W_{\psi, \tau}) = \sum_{c_1,c_2,c_3\in C} \psi(c_1)\tau((c_1,c_2))\tau((c_2, c_3))
\end{align*}
However, by pushing down the summation over $c_1$, we can produce an expression which requires $O(|C|^2)$ time to evaluate.
\begin{align*}
        f(c_2) &= \sum_{c_1\in C} \psi(c_1)\tau((c_1,c_2))\\
        \Phi(Q, F, W_{\psi, \tau}) &= \sum_{c_2,c_3\in C} \tau((c_2, c_3)) f(c_2)
\end{align*}
This version materializes a a vector of intermediate values and then uses that vector in the second line. Doing this allows us to avoid performing an unnecessary summation over 3 variables at once. 

More generally, we can apply this strategy by choosing a variable order and at each step summing out the next variable in the order. The efficacy depends on the maximum number of variables present in any intermediate product. If an intermediate product involves $k$ variables, then we need to compute a relation of size $|C|^k$ which dominates the runtime. We defer the details of the proof to the technical report \cite{tech-report}, but this intuition can be formalized as follows using the theory of tree decompositions and treewidth,
\begin{thm}
\label{thm:inference-tw}
Given a query graph, $Q$, a lifted graph, $F$, and a decomposable estimator $W$, $\Phi(Q, F, W_{\psi, \tau})$ can be computed in time $O(|C|^{\textit{tw}(Q) + 1})$ where $\textit{tw}(Q)$ is the treewidth of $Q$.
\end{thm}

\begin{table*}[]
\small
\caption{Estimator Failure Rates per dataset. Only \textsc{Color} and Characteristic Sets (\texttt{cset}) succeed on all queries. Later we will see \textsc{Color} outperforms \texttt{cset}'s accuracy substantially.}
\begin{tabular}{lp{0.3in}p{0.3in}p{0.3in}p{0.3in}p{0.3in}p{0.3in}p{0.3in}p{0.3in}p{0.3in}p{0.3in}p{0.3in}} \toprule
\textbf{Dataset\textbackslash{}Method} 
& \textbf{cs} & \textbf{wj} & \textbf{jsub} & \textbf{impr} & \textbf{cset} & \textbf{alley} & \textbf{alley\tiny{TPI}} & \textbf{LSS} & \textbf{BSK++} & \textbf{sumrdf} & \textbf{\textsc{Color}} \\
\midrule \textbf{human}  & 0.67 & \textbf{0.00} & 0.22 & 0.63 & \textbf{0.00} & \textbf{0.00} & \textbf{0.00} & \textbf{0.00} & \textbf{0.00} & \textbf{0.00} & \textbf{0.00}\\ 
\textbf{aids}  & 0.69 & 0.07 & 0.14 & 0.28 & \textbf{0.00} & 0.04 & 0.01 & \textbf{0.00} & 0.02 & 0.39 & \textbf{0.00}\\
\textbf{lubm80}  & 0.83 & 0.17 & 0.67 & 0.67 & \textbf{0.00} & \textbf{0.00} & \textbf{0.00} & \textbf{0.00} & 
\textbf{0.00} & \textbf{0.00} & \textbf{0.00}\\
\textbf{yeast}  & 1.00 & 0.97 & 0.97 & 0.11 & \textbf{0.00} & 0.63 & 0.60 & \textbf{0.00} & 0.63 & 0.88 & \textbf{0.00}\\
\textbf{dblp}  & 1.00 & 0.99 & 0.94 & 0.15 & \textbf{0.00} & 0.14 & 0.14 & \textbf{0.00} & 0.70 & 0.85 & \textbf{0.00}\\
\textbf{youtube}  & 0.99 & 0.93 & 1.00 & 0.22 & \textbf{0.00} & 0.10 & 0.05 & \textbf{0.00} & 0.63 & 0.78 & \textbf{0.00}\\ 
\textbf{eu2005}  & 0.95 & 0.90 & 0.91 & 0.55 & \textbf{0.00} & \textbf{0.00} & \textbf{0.00} & \textbf{0.00} & 0.22 & 0.44 & \textbf{0.00}\\
\textbf{patents}  & 0.98 & 0.88 & 0.98 & 0.08 & \textbf{0.00} & 0.13 & 0.13 & \textbf{NA} & 0.67 & 0.79 & \textbf{0.00}\\ \bottomrule
\end{tabular}
\label{tbl:estimator-failure}
\end{table*}
Of course, this relies on finding a good ordering of the query vertices, and finding the optimal one is naively NP-Hard with respect to the size of $Q$. Fortunately, there are very effective heuristics for identifying good tree decompositions, and we apply these in our system, using the min-fill heuristic \cite{DBLP:conf/cp/RollonL11}. To accommodate our sampling techniques, we restrict these tree decompositions to path decompositions and get results relative to the pathwidth.

\subsection{Sampling Techniques}
\label{subsec:sampling-techniques}
In real systems, cardinality estimation needs to be extremely fast and consistent because its overhead is seen by every query. While partial aggregation speeds up inference for simple query graphs with low treewidth, larger and denser query graphs still pose a problem. To avoid this, we use a sampling procedure that ensures linear inference w.r.t query size (see Fig. \ref{fig:partial-agg-youtube}). This is similar to a weighted version of the WanderJoin algorithm from \cite{DBLP:journals/tods/LiWYZ19}. We apply a Thompson-Horowitz estimator to random paths within the lifted graph. However, we adjust the method in two important ways: 1) we incorporate the sampling into the aggregation framework from Sec. \ref{subsec:partial-aggregation} 2) we apply importance sampling to account for the fact that different paths within the lifted graphs contribute more or less to the cardinality estimate.

\paragraph{Sampling During Aggregation}
At each step, we process a vertex of the query graph and materialize an intermediate result consisting of partial colorings and their weights. After materialization, we apply sampling in order to reduce the amount of partial colorings that we extend in the next step. By doing this at each step, we can maintain a constant number of partial colorings at all times and ensure a linear runtime w.r.t. the size of the query graph. Because we still apply aggregation, we reduce the amount of sampling required.

\paragraph{Importance Sampling} 
This is a classical technique for approximating the value of an integral, and we adapt it here by noting that our summation in \ref{eq:optimization:faq_version} is a discrete integral over a product. The core idea is to sample points of the integrand which contribute more heavily to the result with higher probability in order to reduce the variance. Because determining the contribution of a partial coloring to the final result is challenging, we approximate this contribution via its partial count. This assumes that a high partial count leads to a high contribution to the final sum. To keep our estimator unbiased, we apply Thompson-Horowitz estimation and scale each sampled partial count by the inverse of its selection probability. Finally, we scale the total sampled weight to set it to the total weight prior to sampling.

\subsection{Handling Updates}
\label{subsec:updates}
Updates pose a challenge to summary-based estimators because the statistics which they collect become stale over time as updates are applied to the database. Traditional estimators simply rebuild the summary intermittently\cite{ DBLP:journals/dr/Haas99a}. This leads to severe decreases in accuracy under even modest updates because the estimator is entirely "blind" to them. On the other hand, recalculating the summary before each query is very costly. In this section, we demonstrate how COLOR supports a middle ground approach that applies fast, basic updates to the lifted graph, allowing it to maximize the time between full rebuilds.

First, we formally define updates in our setting,
\begin{definition}
    \label{def:update}
    Given a data graph G, an update $\theta$ can either add an edge  between existing vertices or a new vertex with attributes $A$:
    \begin{itemize}
        \item $\theta_V = (v, A)$ where $v \in G_V, A\in\mathbb{A}$
        \item $\theta_E = (v_1, v_2, A)$ where $v_1 \in G_V, v_2 \in G_V, A\in\mathbb{A}$
    \end{itemize}
\end{definition}

This definition allows for adding a single edge or vertex to the graph at a time. We then define a summary update function to incorporate these updates without accessing $G$.
\begin{definition}
    \label{def:summary-update}
    Given a data graph $G$, a lifted graph $F = (G_F, \psi, \tau)$, and update $\theta$, we define the summary update function as follows where $F'$ is the updated lifted graph,
    \begin{align*}
         F' = \delta (F, \theta)
    \end{align*}
    Depending on the type of $\theta$, the functionality of $\delta$ can change:
    \begin{align*}
    \delta &= \begin{cases}
    \delta_V, & \text{$\theta \in \theta_V$} \\
    \delta_E, & \text{$\theta \in \theta_E$} 
    \end{cases}
\end{align*}
\end{definition}

Depending on the estimator, the correct definition of $\delta$ will change. Here, we focus on the average degree estimator.

\paragraph{Vertex Updates} The vertex update function $\delta_V$ affects the stored edge statistics $\tau$ and color sizes $\psi$ in the lifted graph. Because a new vertex has no edges, we have no knowledge about which color it should be placed once its edges are added. Conservatively, we simply add it to the largest existing color which dilutes the impact on the average degree. In this way, we preserve the high quality information in other colors while the largest color gracefully degrades to a traditional estimator as in Def. \ref{def:traditional-estimator}.  After choosing the color for the new vertex, we adjust $\psi(c)$ by incrementing its value by one, and we scale down $\tau$ to account for the new vertex.
\paragraph{Edge Updates} Given an update $\theta_E=(v_1,v_2,A)$, the edge update function $\delta_E$ marginally increases $\tau$ for the combination of attributes and colors in the edge update. However, the edge update doesn't directly contain the colors of $v_1$ and $v_2$ or the attributes of $v_2$.  To retrieve the colors, $c_1$ and $c_2$, associated with $v_1$ and $v_2$, we look up their values in $\pi$ which we store compactly (and approximately) as a series of cuckoo filters. To calculate the attributes of $v_2$, we leverage statistics about the attribute distribution in $c_2$ to sample a set of attributes.
\paragraph{Path Closure Probabilities} The lifted graph contains statistics about the cycle-closing probability for existing nodes and edges, but additions to the graph change this probability. To account for this, we make an adjustment to the $\omega^\circ_{CCP}$ function. We calculate the probability that either the path was originally closed, $\gamma(c_1,c_2,D)$, or is closed by an update edge. For a set of edge updates, $\mathbb{S}_{\theta E}$:
\begin{align*}
    \gamma'(c_1, c_2, D) &= 1- (1 - \gamma(c_1,c_2,D))(1-\frac{|\mathbb{S}_{\theta E}|}{|V_F|^2})
\end{align*}

\paragraph{Deletions} To handle deletions, COLOR simply performs the inverse of the update logic.
\begin{figure*}[]
    \centering
    \includegraphics[width=.9\textwidth]{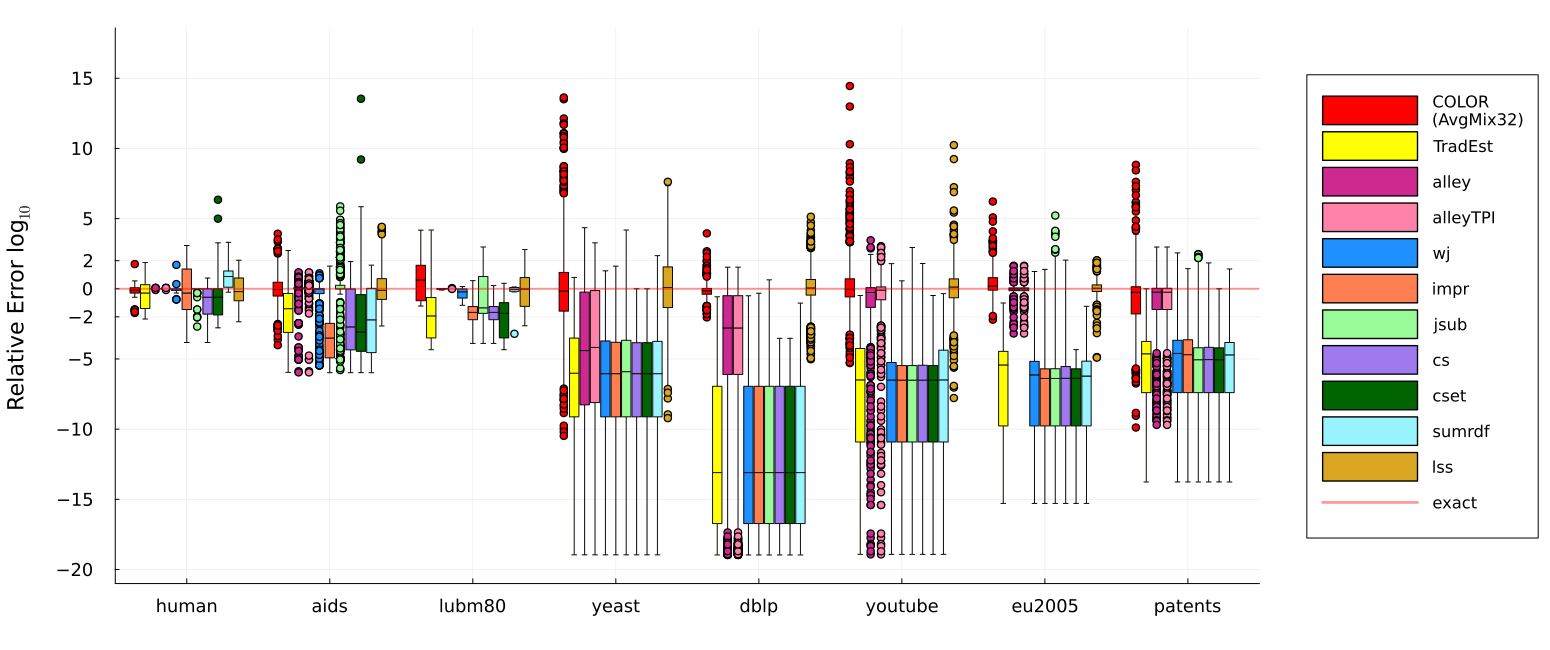}
    \caption{Relative Error by Estimator}
    \label{fig:relative-error}
\end{figure*}

\begin{figure*}[]
    \centering
    \includegraphics[width=.9\textwidth]{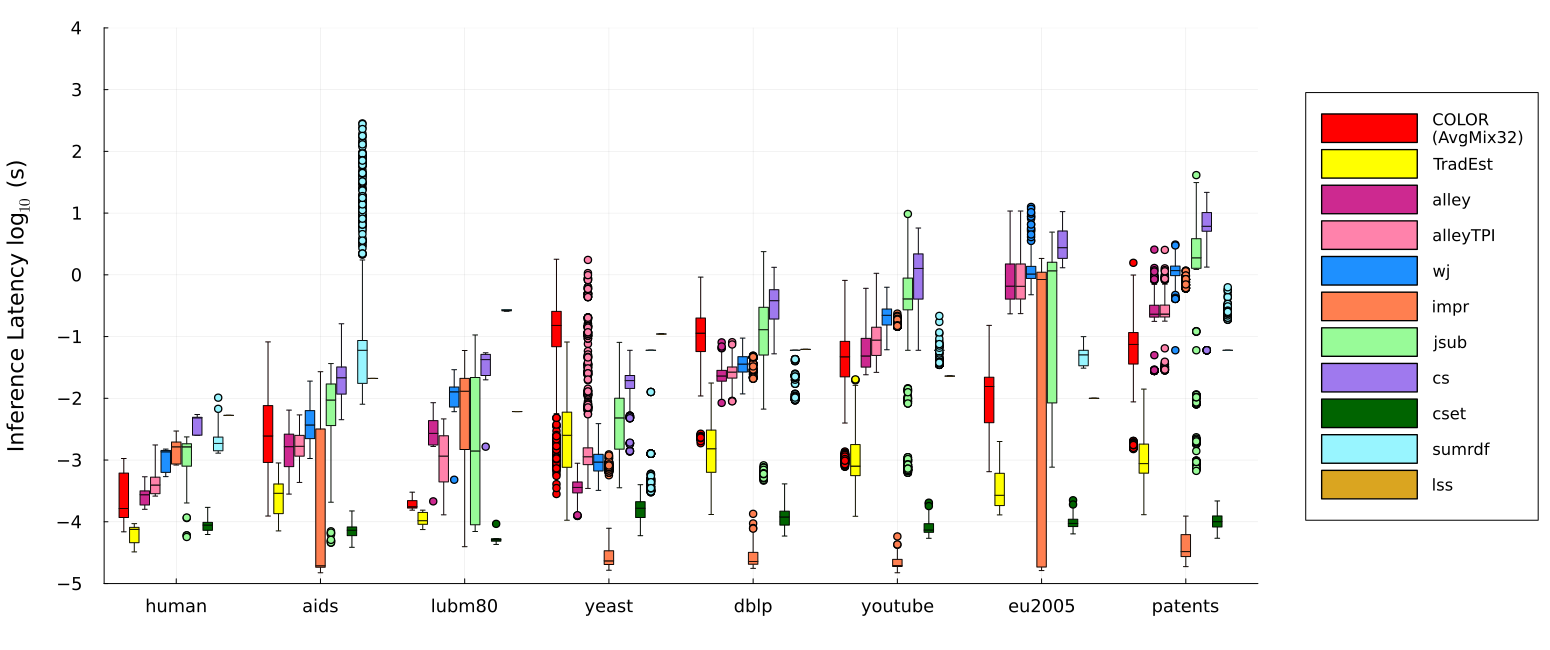}
    \caption{Inference Time by Estimator}
    \label{fig:inference-time}
\end{figure*}

\begin{figure}
    \centering    \includegraphics[width=0.35\textwidth]{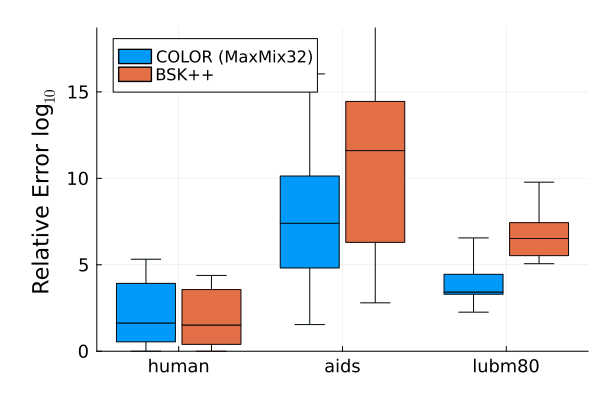}
    \caption{Relative Error by Cardinality Bound Method}
    \label{fig:bounds-relative-error}
\end{figure}

\begin{figure}
    \centering    \includegraphics[width=0.35\textwidth]{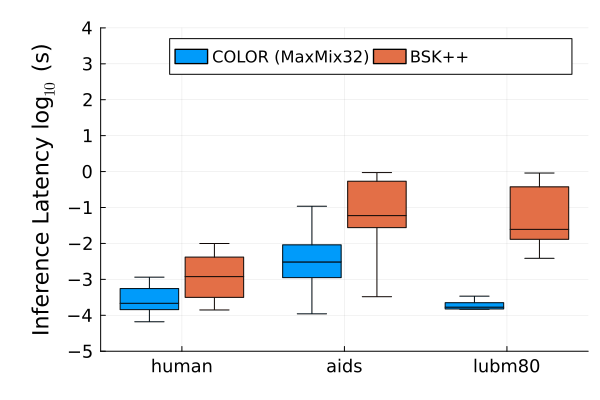}
    \caption{Inference Time by Cardinality Bound Method}
    \label{fig:bounds-inference-time}
\end{figure}

\section{Evaluation}
\label{sec:evaluation}
In this section, we provide a detailed experimental analysis of our framework on eight benchmark datasets and against nine competitive baselines. Compared to existing methods, COLOR exhibits competitive accuracy, speed, and scalability. We also demonstrate the significance of our performance optimizations for building and maintaining graph summaries. Overall, we show that COLOR:
\begin{enumerate}
    \item Never experiences estimation failure on any query across all workloads.
    \item Has a median error $\mathbf{<10}$ across all workloads and is up to $\mathbf{10^3\times}$ more accurate than competing methods.
    \item Produced up to $\mathbf{10^7\times}$ tighter bounds when using $\tau_{max}$ than BoundSketch while being many orders of magnitude faster.
    \item Requires up to $\mathbf{80-800\times}$ less space than competing methods and is up to $\mathbf{10-100\times}$ faster to construct.
    \item Handles updates gracefully with only $\mathbf{3\times}$ worse median error when half the graph is updated. 
\end{enumerate}

\begin{table}[]
\small
\caption{Experimental Datasets. $|\ell_V|$ and $|\ell_E|$ are the number of unique vertex and edge labels.}
\begin{tabular}{lcccl}
\toprule
\textbf{Dataset} & \textbf{|V|} & \textbf{|E|} & \textbf{|$\ell_V$|} & \textbf{|$\ell_E$|} \\ \midrule
human            & 4674         & 86282        & 89                  & 1                   \\ 
aids             & 254000       & 548000       & 50                  & 4                   \\ 
lubm80           & 2.6M         & 12.3M        & 35                  & 35                  \\
yeast            & 3112         & 12519        & 71                  & 1                   \\
dblp             & 317080       & 1M           & 15                  & 1                   \\
youtube          & 1.1M         & 3M           & 25                  & 1                   \\
eu2005           & 862664       & 16M          & 40                  & 1                   \\
patents          & 3.8M         & 16.5M        & 20                  & 1                   \\ \bottomrule
\end{tabular}
\label{tbl:datasets}
\end{table}

\textit{Datasets \& Workload.} We consider datasets from \cite{DBLP:conf/sigmod/ParkKBKHH20} and \cite{sun2020memory} for our analysis. These datasets come from a variety of domains. Those from \cite{sun2020memory} are undirected graphs whose queries are larger and vary in density. Those from \cite{DBLP:conf/sigmod/ParkKBKHH20} are directed graphs whose queries are smaller and vary in shape. We adapt undirected workloads to directed methods by including reverse edges in the data graph but not in the query graphs. Both of these sources include label predicates in their queries with the former using both edge and vertex labels and the latter using only vertex labels. Table \ref{tbl:datasets} shows their different characteristics where $|\ell_V|$ and $|\ell_E|$ are the number of edge and vertex labels.

\textit{Comparison Methods.} For comparison, we use a superset of the methods considered in \cite{DBLP:conf/sigmod/ParkKBKHH20} and additionally apply them to the larger, more complex datasets from \cite{sun2020memory}. These methods include: 1) Correlated Sampling (CS) \cite{DBLP:journals/pvldb/VengerovMZC15} 2) Characteristic Sets (CSet)\cite{DBLP:conf/icde/NeumannM11} 3) Wander Join (WJ) \cite{DBLP:journals/tods/LiWYZ19} 4) Alley (alley) and alleyTPI \cite{DBLP:conf/sigmod/KimKFH21}  5) Join Sampling with Upper Bounds (JSUB), an adaptation of \cite{DBLP:conf/sigmod/ZhaoC0HY18} 6) Bound Sketch (BSK) \cite{DBLP:conf/sigmod/CaiBS19} which corresponds to a hash-based coloring and using the max degree estimator. We further apply our partial aggregation optimization, and we call this improved version BSK++. 7) IMPR \cite{DBLP:journals/tkdd/ChenL18} and 8) SumRDF \cite{DBLP:conf/www/StefanoniMK18} \rthree{9) Learned Sketch for Subgraph Counting (LSS) \cite{DBLP:journals/vldb/ZhaoYLZR23}, a query-driven deep learning approach. Due to hardware compatibility issues, this method was run on a different machine than the others with an Intel(R) Core(TM) i7-9750H CPU @ 2.60GHz CPU. This machine had 32GB of memory, resulting in an OOM error for the patents dataset which we treat as NA rather than estimator failure.} 10) We also include a traditional independence-based estimator (IndEst) corresponding to Def. \ref{def:traditional-estimator}. For the sampling based estimators, we apply the default sampling ratios from \cite{DBLP:conf/sigmod/ParkKBKHH20} and \cite{DBLP:conf/sigmod/KimKFH21} (i.e. .03 for all methods except for Alley which uses .001). AlleyTPI uses a maximum pattern length (MAX\_L) and a maximum number of stored label groups (NUM\_GROUPS) when building its index, with default values of MAX\_L=4 or MAX\_L=5 depending on the dataset, and NUM\_GROUPS=32. To prevent excessive index build times on AlleyTPI ($>12$ hours) for eu2005, dblp, and patents, we used MAX\_L=4. We decreased NUM\_GROUPS to 16 for eu2005 and dblp and 8 for patents. We adjusted NUM\_GROUPS more than MAX\_L because small adjustments to the maximum pattern length greatly decrease the domain of patterns stored by the index \cite{DBLP:conf/sigmod/KimKFH21}. \rthree{As in the original work, we get estimates for LSS via a 5-fold cross-validation where 4/5ths of the queries are trained on and 1/5 is estimated during each fold. Lastly, recently, there has been work on the isomorphism variant of subgraph cardinality estimation from both the deep learning and sampling perspective\cite{DBLP:journals/pvldb/ShinSPH24,DBLP:conf/sigmod/WangH000022}. Unfortunately, this difference in problem setting makes them incomparable with the other methods considered here, so we exclude them from our evaluation.}

Additionally, we experiment with several instantiations of our framework which use the following naming convention; first, we note the kind of degree statistic (Min/Avg/Max), then we describe the coloring scheme, e.g. $Q64$ as 64 colors from the quasi-stable coloring method. The mixed coloring scheme,  $Mix32$, that we use as the default involves 8 divisions from degree coloring, quasi-stable coloring, neighbor labels coloring, and node label coloring, in that order. Unless otherwise noted, we use 500 samples during inference and keep track of cycle probabilities for cycles up to length 6.

\textit{Experimental Setup.} To reduce noise, we repeat all inference results 3 times and report the median inference time. We do not do this for our cardinality estimates because this would unfairly reduce the impact of our sampling approach. These experiments are run on a server with an Intel(R) Xeon(R) CPU E7-4890 v2 @ 2.80GHz CPU, and all summary building and inference is done using a single thread, unless stated otherwise. The reference implementation is available at: \url{https://github.com/uwdb/color} 


\subsection{Estimator Failure}
\label{subsec:estimator-failure}
There are two ways that an estimator can fail to provide meaningful results: 1) time outs, which we define as taking longer than 1 minute to report a result 2) sampling failure, not finding any qualifying samples. In Table \ref{tbl:estimator-failure}, we show the proportion of queries that result in estimation failure for each dataset and technique. Simpler sampling-based methods (CS, WJ, JSUB, IMPR) face estimation failure even on the smaller query workloads and fail to find any samples most queries on the larger workloads. Alley achieves much higher success rates due to its sophisticated sampling approach but still fails a significant portion of the time on four datasets. Summary-based methods (BSK++ and SumRDF), on the other hand, time out on over half of queries for four datasets. In contrast, because our approach applies sampling to a highly dense lifted graph, we never experience sampling failure on any query, across all workloads. As the data graph contains hundreds of thousands of edges or more, which are mapped into a lifted graph with at most 32 nodes (colors), the probability that any two colors have an edge connecting them is high.

\subsection{Accuracy} 
\label{subsec:accuracy}
In Fig. \ref{fig:relative-error}, we show the relative error of various methods and workloads\footnote{In these graphs, outliers (>2 std. deviations) are shown as points. The inner box shows quartiles, and the whiskers are the max/min non-outlier values. Further, sample failure is treated as an estimate of 1.}. Relative error is defined as the ratio of the estimate to the true cardinality. Scores greater than one indicate an overestimate, those under an underestimate. Across workloads, our method, AvgMix32, is unbiased and scales well to larger, more cyclic workloads (yeast, dblp, youtube, eu2005, patents). It even achieves a median error of less than 2 on human, aids, dblp, and eu2005. We also reproduce the high accuracy of WanderJoin and Alley on the G-Care datasets. However, we find that all methods from \cite{DBLP:conf/sigmod/ParkKBKHH20} fail to scale to the larger more cyclic workloads. In particular, we reproduce the finding in \cite{DBLP:conf/sigmod/KimKFH21} that WanderJoin, IMPR, and JSUB overwhelmingly fail to find a positive sample in a reasonable time on these datasets, \rthree{and that LSS achieves reasonable accuracy on a variety of workloads.} Further, SumRDF times out on all larger queries.

\paragraph{Cardinality Bounds} When comparing the cardinality bounding methods, BSK and MaxQ64, we find that applying a mixture of coloring methods rather than hash coloring produces up to $10^6$ times lower error. Notably, because we apply sampling to this method, we guarantee a linear runtime for all queries in exchange for a less principled cardinality bound. However, across all workloads, this never results in significant underestimation.

\begin{figure}
    \centering
    \includegraphics[width=0.4\textwidth]{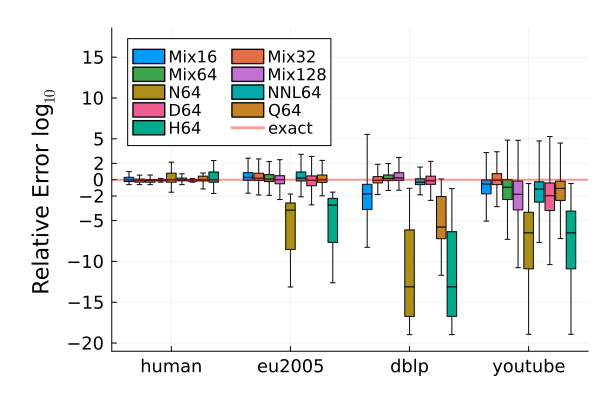}
    \caption{Relative Error by Coloring Method}
    \label{fig:colorings-error}
\end{figure}

\paragraph{Coloring Methods} In Fig. \ref{fig:colorings-error}, we examine the effect of choosing different colorings on the accuracy of the average degree estimator. The hash coloring performs the worst across all benchmarks which is expected because it does not take the labels or graph topology into account. On the other hand, the quasi-stable coloring algorithm from \cite{DBLP:journals/pvldb/KayaliS22} works quite well on most datasets with the exception of dblp because it does not account for the distribution of labels. Overall, the mixed coloring performs well across datasets because it can supplement the topological colorings with a label-based colorings, accounting for both sources of error. 

In this figure, we also vary the number of colors that we use for the lifted graph. Interestingly, increasing the number of colors kept does not straightforwardly improve accuracy. This is because we always use 500 samples during inference. As the lifted graph gets larger, the sampling procedure has a larger impact. Given this, we find that the optimal coloring uses either 32 or 64 colors.

\subsection{Inference Latency}
\label{subsec:inference-latency}
Fig. \ref{fig:inference-time} shows the distribution of inference latencies for each method across workloads. We can see that the inference latency of COLOR lies in the middle of the competing methods across workloads. On the smaller, less cyclic queries of human, it achieves a median latency of around $10^{-4}$ seconds due to partial aggregation, and on the more complex queries of patents it has a median latency of $\sim .05$ seconds via sampling.
Similar to prior work, we select a one minute timeout because cardinality estimators may be called many times during the query planning phase~\cite{DBLP:journals/pvldb/WangQWWZ21, DBLP:conf/sigmod/KimKFH21, DBLP:journals/vldb/ZhaoYLZR23, park2020g}. 

Compared to other graph summarization methods, SumRDF and BSK++, our framework scales far better to larger queries. The former methods timeout on queries of even moderate size as they consider the exponential number of potential colorings of the query. This occurs even when using partial aggregation (as in BSK++), demonstrating the necessity of sampling to achieve consistent latencies.

\subsection{Statistics Size \& Build Time}
\label{subsec:stat-size-and-build-time}
\begin{figure}
    \centering
    \includegraphics[width=.45\textwidth]{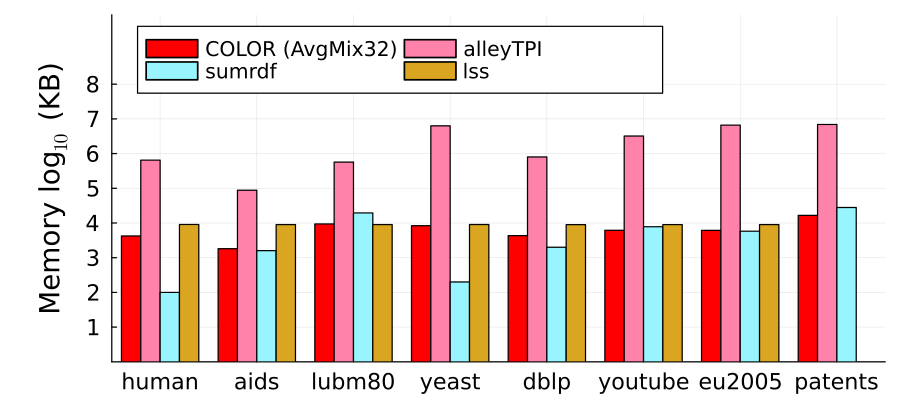}
    \caption{\rthree{Statistics Size}}
    \label{fig:statistics-size}
\end{figure}

Graph summarization approaches allow for a smooth tradeoff between accuracy and size/build time; a more granular summary of the graph will take more space but more accurately capture the structure of the data graph. Even a compact summary ($<20$MB) can be highly accurate as shown by Fig. \ref{fig:statistics-size}. This comes from the fact that we store our summary sparsely. If two colors do not share an edge with a particular attribute, then we don't explicitly store any degree statistic about this combination. A better coloring can actually result in a more compact summary because many of these combinations won't occur if the nodes have been properly partitioned. On Human and Lubm80, AlleyTPI's pattern index requires 600 and 300 MB, respectively. \rthree{LSS takes a considerable time to train across all datasets. This is the average training time over all folds of the cross validation, and, crucially, it does not include the time necessary to collect the training data (i.e. to run the queries and calculate a true cardinality).}

\begin{figure}
    \centering
    \includegraphics[width=.35\textwidth]{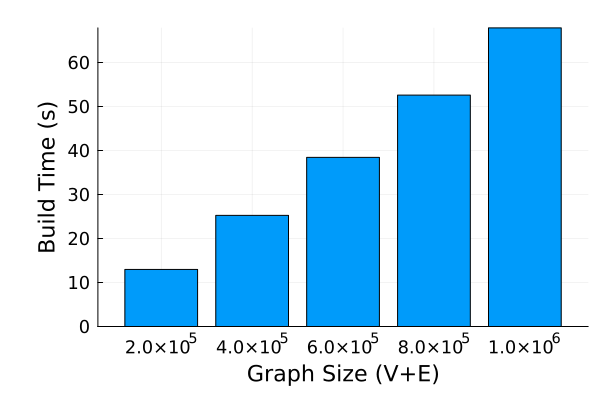}
    \caption{Construction Scaling}
    \label{fig:const-scaling}
\end{figure}

With respect to build time, our summary construction scales linearly in the size of the data graph. For smaller graphs like human, aids, and yeast, summary construction takes less than $20$ seconds, and it smoothly increases as the data graph gets larger. In Fig. \ref{fig:const-scaling}, we confirm this by generating erdos-renyi graphs of varying sizes and recording the average time to build the lifted graph over 20 trials.

\begin{figure}
    \centering
    \includegraphics[width=.45\textwidth]{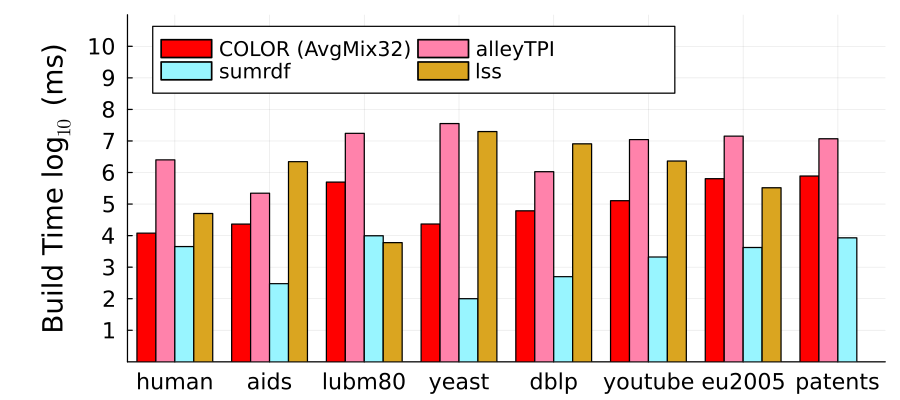}
    \caption{\rthree{Build Time}}
    \label{fig:build-time}
\end{figure}

\subsection{Updates}
\label{subsec:updates-evaluation}
\begin{figure} 
    \centering
    \includegraphics[width=0.38\textwidth]{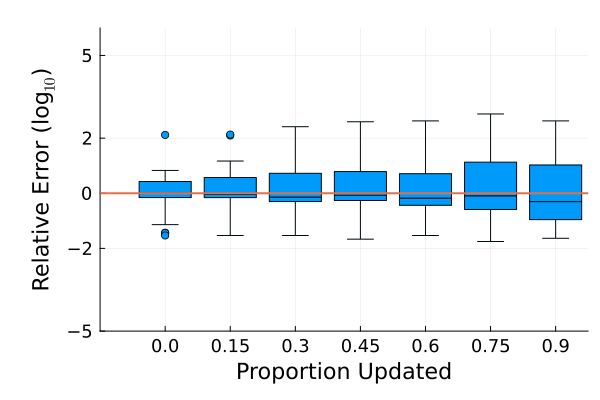}
    \caption{Relative Error vs Proportion Updated (Human)}
    \label{fig:proportion-updated-error}
\end{figure}

In Fig. \ref{fig:proportion-updated-error}, we evaluate the effectiveness of our method for handling updates (Sec. \ref{subsec:updates}). To do this, we randomly partition the edges of the human dataset into an initial data graph and an ensuing set of updates, and we construct a lifted graph using the former then update it by adding one edge/vertex at a time based on the latter. Intuitively, when more of the graph is provided at the beginning, the lifted graph will be more accurate because it can take advantage of that knowledge when coloring the graph. We find that, as more of the lifted graph is updated, the accuracy remains very consistent and degrades to the traditional independence estimator. So, a full rebuilding of the lifted graph can occur very infrequently and be amortized over many updates. Due to the small size of the lifted graph, each update is very fast. The latency for vertex updates was roughly 0.4 ms while for edge updates it was roughly 0.1 ms.

\subsection{Micro-Benchmarks}
\label{subsec:micro-benchmarks}

\begin{figure}
    \centering
    \includegraphics[width=0.38\textwidth]{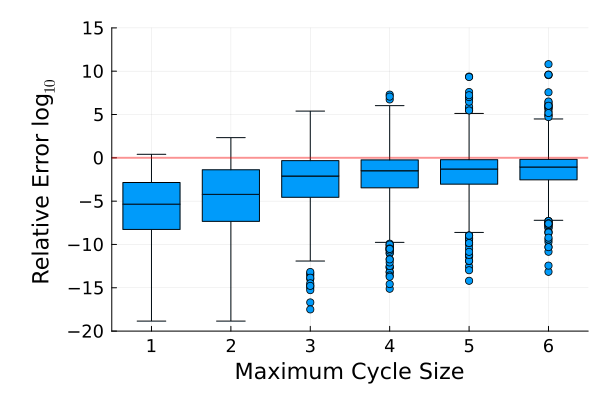}
    \caption{Relative Error vs Max Cycle Length Stored (Youtube)}
    \label{fig:cycle-sizes-error}
\end{figure}

\paragraph{Path Closure Probabilities} To show the importance of tracking cycle probabilities, we show the relative error as we vary the size of cycles whose probabilities we track in Fig. \ref{fig:cycle-sizes-error}. At size 1, we do not track any cycle closure probabilities. So, when we close a cycle in the query graph, we scale down the estimate based on the uniform probability of an edge existing, i.e. $|E|/|V|^2$. When we begin to store larger cycle sizes, we quickly see the error decrease, and underestimation, in particular, is significantly reduced. These results validate the necessity of handling cycle closure with a more complex method than the standard independence estimator.

\begin{figure}
    \centering
    \includegraphics[width=0.38\textwidth]{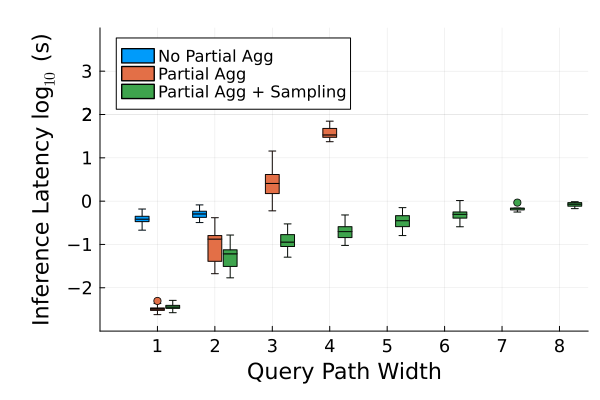}
    \caption{Inference Time vs Query Pathwidth (Youtube)}
    \label{fig:partial-agg-youtube}
\end{figure}

\paragraph{Partial Aggregation} Fig. \ref{fig:partial-agg-youtube} demonstrates the effects of partial agg. and sampling. Using the Youtube dataset, we study the effect of partial agg. and sampling on inference latency across queries with different pathwidths. Recall that pathwidth is a measure of cyclicity where pathwidth 1 is acyclic and higher pathwidth queries are denser. Without partial agg., estimation times out (>1 min) when pathwidth exceeds 2. Higher pathwidth queries are larger and the naive approach is exponential w.r.t. the query size. Using only partial agg., estimation times out when pathwidths exceed 4. Lastly, when sampling is applied, the inference latency becomes linear in the size of the query and unrelated to the pathwidth. The results demonstrate that partial agg. achieves speedups without affecting accuracy and sampling can achieve consistent, fast inference.

\begin{figure}
    \centering
    \includegraphics[width=0.38\textwidth]{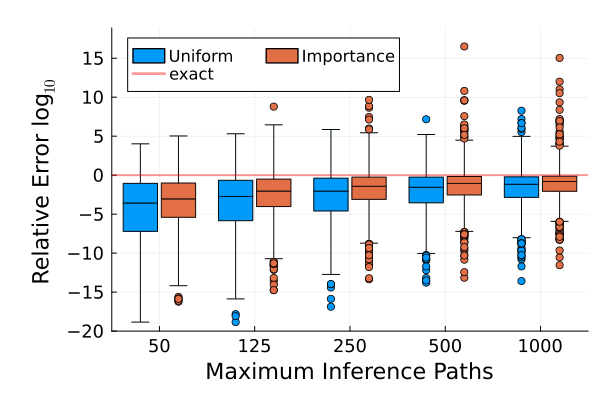}
    \caption{Relative Error vs Samples (Youtube)}
    \label{fig:inference-samples-error}
\end{figure}

\paragraph{Inference Sampling} In Fig. \ref{fig:inference-samples-error}, we vary sample size to demonstrate the efficacy of our sampling method. Using a very small sample results in significant underestimation, but even a moderate number of samples quickly converges to the accuracy of a large number of samples. Further, this figure shows that importance sampling speeds up this convergence significantly over a naive uniform sample. For example, the performance at $250$ samples for importance sampling is roughly equal to the accuracy of uniform sampling at $1000$ samples.  

\section{Conclusion}
We develop COLOR, a framework for producing lifted graph summaries from colorings. We define inference over the lifted graph for acyclic and cyclic queries, developing optimizations to accelerate estimation. We empirically validate COLOR's superior performance on eight benchmarks and compare to state-of-the-art methods. COLOR is up to $10^3 \times$ more accurate than the baselines and stands out for never experiencing estimation failure. It gracefully handles updates, degrading to only $3\times$ worse error when half of the data is replaced.



\begin{acks}

This work was partially supported by NSF-BSF 2109922, NSF IIS 2314527, and a gift from Amazon through the UW Amazon Science Hub.

\end{acks}


\bibliographystyle{ACM-Reference-Format}
\bibliography{sample}

\end{document}